\documentstyle[12pt]{article}

\def\noi{\noindent}

\makeatletter
\renewcommand{\section}{\@startsection{section}{1}{0pt}%
        {-3.5ex plus -1ex minus -.2ex}{2.3ex plus .2ex}%
        {\large\bf\protect\raggedright}}

\renewcommand{\subsection}{\@startsection{subsection}{2}{0pt}%
        {-3ex plus -1ex minus -.2ex}{1.4ex plus .2ex}%
        {\normalsize\bf\protect\raggedright}}

\renewcommand{\thesubsubsection}%
        {\arabic{section}.\arabic{subsection}.\arabic{subsubsection}.}

\renewcommand{\@oddhead}{\raisebox{0pt}[\headheight][0pt]{%
   \vbox{\hbox to\textwidth{\rightmark \hfil \rm \thepage \strut}\hrule}}}
\renewcommand{\@evenhead}{\raisebox{0pt}[\headheight][0pt]{%
   \vbox{\hbox to\textwidth{\thepage \hfil \leftmark \strut}\hrule}}}
\newcommand{\heads}[2]{\markboth{\protect\small\it #1}{\protect\small\it #2}}
\newcommand{\Acknow}[1]{\subsection*{Acknowledgement} #1}
\makeatother

\newcommand{\Title}[1]{\noi {\Large #1} \\}
\newcommand{\Author}[2]{\noi{\large\bf #1}\\[2ex]\noindent{\it #2}\\}

\newcommand{\Abstract}[1]{\vskip 2mm \begin{center}
        \parbox{16.4cm}{\small\noi #1} \end{center}\medskip}

\newcommand{\foom}[1]{\protect\footnotemark[#1]}

\newcommand{\email}[2]{\footnotetext[#1]{e-mail: #2}}

\def\nqq{\hspace*{-2em}}

\def\cm{\hspace*{1cm}}

\def\ten#1{\mbox{$\cdot 10^{#1}$}}

\def\lal{&&\nqq {}}

\def\beq{\begin{equation}}
\def\eeq{\end{equation}}
\def\bear{\begin{eqnarray}}
\def\bearr{\begin{eqnarray} \lal}
\def\ear{\end{eqnarray}}
\def\earn{\nonumber \end{eqnarray}}
\def\nn{\nonumber\\ {}}

\def\yy{\\[5pt] {}}

\def\d{\partial}

\def\const{{\rm const}}

\def\Gun{\mbox{m$^3$ kg$^{-1}$ s$^{-2}$}}
\def\year{{\rm year}}
\def\R{{mathbb R}}

\heads
{Vitaly N. Melnikov}
{Gravity as a Key Problem of the Millenium}

\begin{document}

\Title
{GRAVITY AS A KEY PROBLEM OF THE MILLENIUM}

\Author{Vitaly N. Melnikov}
{Institute of Gravitation and Cosmology, Peoples' Friendship
	University of Russia, 6 Miklukho-Maklaya St., Moscow 117198, and \\
Center for Gravitation and Fundamental Metrology,
	VNIIMS, 3-1 M. Ulyanovoy St., Moscow 117313, Russia\\
and Depto.de Fisica, CINVESTAV, Apartado Postal 14-740, Mexico 07360,
D.F.\foom 1}
\email 1 {melnikov@rgs.phys.msu.su, melnikov@fis.cinvestav.mx}

\Abstract
{Gravitation as a fundamental interaction that governs all phenomena at
large and very small scales, but still not well understood at a quantum
level, is a missing cardinal link to unification of all
physical interactions. Problems of the absolute G measurements and its
possible time and range variations are reflections of the unification
problem. Integrable multidimensional models of gravitation and cosmology
make up one of the proper approaches to study basic issues and strong
field objects, the Early Universe and black hole physics in particular.
 The choice, nature, classification and precision of determination of
fundamental physical constants are described. The problem of their temporal
variations is also discussed, temporal and range variations of $G$ in
particular. A need for further absolute measurements of
$G$, its possible range and time variations is pointed out. The novel
multipurpose space project SEE, aimed for measuring G and its stability
in space and time 3-4 orders better than at present, may answer many
important questions posed by gravitation, cosmology and unified theories.}

\section{Introduction}

The second half of the 20th century in the field of gravitation was
devoted mainly to theoretical study and experimental verification of
general relativity and alternative theories of gravitation with a strong
stress on relations between macro and microworld fenomena or, in other
words, between classical gravitation and quantum physics. Very intensive
investigations in these fields were done in Russia by M.A.Markov,
K.P.Staniukovich, Ya.B.Zeldovich, A.D.Sakharov and their colleagues
starting from mid 60's. As a motivation there were: singularities in
cosmology and black hole physics, role of gravity at large and very small
(planckian) scales, attempts to create a quantum theory of gravity as for
other physical fields, problem of possible variations of fundamental
physical constants etc. A lot of work was done along such topics as
\cite{3}:

- particle-like solutions with a gravitational field,

- quantum theory of fields in a classical gravitational background,

- quantum cosmology with fields like a scalar one,

- self-consistent treatment of quantum effects in cosmology,

- development of alternative theories of gravitation: scalar-tensor,
gauge, with torsion, bimetric etc.

As all attempts to quantize general relativity in a usual manner failed
and it was proved that it is not renormalizable, it bacame clear that the
promising trend is along the lines of unification of all physical
interactions which started in the 70's. About this time the experimental
investigation of gravity in strong fields and gravitational waves started
giving a powerful speed up in theoretical studies of such objects as
pulsars, black holes, QSO's, AGN's, Early Universe etc., which continues
now.

But nowadays, when we think about the most important lines of future
developments in physics, we may forsee that gravity will be essential not
only by itself, but as a missing cardinal link of some theory, unifying
all existing physical interactions: week, strong and electromagnetic ones.
Even in experimental activities some crucial next generation experiments
verifiing predictions of unified schemes will be important. Among them
are: STEP - testing the corner stone Equivalence Principle, SEE - testing
the inverse square law (or new nonnewtonian interactions), EP, possible
variations of the newtonian constant G with time, absolute value of G with
unprecedented accuracy \cite{15}. Of course, gravitational waves problem,
verification of torsional, rotational (GPB), 2nd order and strong field
effects remain important also.

We may predict as well that thorough study of gravity itself and within
the unified models will give in the next century and millenium even more
applications for our everyday life as electromagnetic theory gave us in
the 20th century after very abstract fundamental investigations of
Faraday, Maxwell, Poincare, Einstein and others, which never dreamed about
such enormous applications of their works.

Other very important feature, which may be envisaged, is an increasing
role of fundamental physics studies, gravitation, cosmology and
astrophysics in particular, in space experiments. Unique microgravity
environments and modern technology outbreak give nearly ideal place for
gravitational experiments which suffer a lot on Earth from its relatively
strong gravitational field and gravitational fields of nearby objects due
to the fact that there is no ways of screening gravity.

In the developement of relativistic gravitation and dynamical cosmology
after A. Einstein and A. Friedmann, we may notice three distinct stages:
first, investigation of models with matter sources in the form of a
perfect fluid, as was originally done by Einstein and Friedmann. Second,
studies of models with sources as different physical fields, starting from
electromagnetic and scalar ones, both in classical and quantum cases (see
\cite{3}). And third, which is really topical now, application of ideas and
results of unified models for treating fundamental problems of cosmology
and black hole physics, especially in high energy regimes.
Multidimensional gravitational models play an essential role in the latter
approach.

The necessity of studying multidimensional models of gravitation and
cosmology \cite{Mel2,Mel} is motivated by several reasons. First, the
main trend of modern physics is the unification of all known fundamental
physical interactions: electromagnetic, weak, strong and gravitational
ones. During the recent decades there has been a significant progress in
unifying weak and electromagnetic interactions, some more modest
achievements in GUT, supersymmetric, string and superstring theories.

Now, theories with membranes, $p$-branes and more vague M- and F-theories
are being created and studied. Having no definite successful theory of
unification now, it is desirable to study the common features of these
theories and their applications to solving basic problems of modern
gravity and cosmology.  Moreover, if we really believe in unified
theories, the early stages of the Universe evolution and black hole physics,
as unique superhigh energy regions, are the most proper and natural arena
for them.

Second, multidimensional gravitational models, as well as scalar-tensor
theories of gravity, are theoretical frameworks for describing
possible temporal and range variations of fundamental physical constants
\cite{3,4,5,6}. These ideas have originated from the earlier papers of E.
Milne (1935) and P. Dirac (1937) on relations between the phenomena of
micro- and macro-worlds, and up till now they are under thorough study
both theoretically and experimentally.

Lastly, applying multidimensional gravitational models to basic problems
of modern cosmology and black hole physics, we hope to find answers to such
long-standing problems as singular or nonsingular initial states, creation
of the Universe, creation of matter and its entropy, acceleration,
cosmological constant, origin of inflation and specific scalar fields
which may be necessary for its realization, isotropization and graceful
exit problems, stability and nature of fundamental constants \cite{4},
possible number of extra dimensions, their stable compactification etc.

Bearing in mind that multidimensional gravitational models are certain
generalizations of general relativity which is tested reliably for weak
fields up to 0.001 and partially in strong fields (binary pulsars), it is
quite natural to inquire about their possible observational or
experimental windows.  From what we already know, among these windows are:

-- possible deviations from the Newton and Coulomb laws, or new
interactions,

-- possible variations of the effective gravitational constant with
a time rate smaller than the Hubble one,

-- possible existence of monopole modes in gravitational waves,

-- different behaviour of strong field objects, such as multidimensional
black holes, wormholes and $p$-branes,

-- standard cosmological tests etc.

Since modern cosmology has already become a unique laboratory for testing
standard unified models of physical interactions at energies that are far
beyond the level of the existing and future man-made accelerators and other
installations on Earth, there exists a possibility of using cosmological
and astrophysical data for discriminating between future unified schemes.

As no accepted unified model exists, in our approach we adopt simple, but
general from the point of view of number of dimensions, models based on
multidimensional Einstein equations with or without sources of different
nature:

-- cosmological constant,

-- perfect and viscous fluids,

-- scalar and electromagnetic fields,

-- their possible interactions,

-- dilaton and moduli fields,

-- fields of antisymmetric forms (related to $p$-branes) etc.

Our program's main objective was and is to obtain exact
self-consistent solutions (integrable models) for these models and then to
analyze them in cosmological, spherically and axially symmetric cases. In
our view this is a natural and most reliable way to study highly nonlinear
systems.  It is done mainly within Riemannian geometry. Some simple
models in integrable Weyl geometry and with torsion were studied as well.

Here we dwell mainly upon some problems of fundamental physical constants,
the gravitational constant in particular, upon the SEE project shortly
(see A.Sanders' paper in this volume)
and exact solutions in the spherically symmetric case,  black hole and
PPN parameters for these
solutions in particular, within a multidimensional gravity.

\section{Fundamental physical constants}

{\bf 2.1.} In any physical theory we meet constants which
characterize the stability properties of different types of matter: of
objects, processes, classes of processes and so on. These constants are
important because they arise independently in different situations and
have the same value, at any rate within accuracies we have gained nowadays.
That is why they are called fundamental physical constants (FPC) \cite{3}.
It is impossible to define strictly this notion.  It is because the
constants, mainly dimensional, are present in definite physical theories.
In the process of scientific progress some theories are replaced by more
general ones with their own constants, some relations between old and new
constants arise.  So, we may talk not about an absolute choice of FPC, but
only about a choice corresponding to the present state of the physical
sciences.

Really, before the creation of the electroweak interaction theory and some
Grand Unification Models, it was considered that this {\em choice} is as
follows:
\beq
c,\ \hbar,\ \alpha,\ G_{F},\ g_s,\ m_p\ ({\rm or}\ m_e),\ G,\ H,\ \rho,\
\Lambda,\ k,\ I,
\eeq
where $\alpha$, $G_F$, $g_s$ and $G$ are constants of electromagnetic,
weak, strong and gravitational interactions, $H$, $\rho$ and $\Lambda$ are
cosmological parameters (the Hubble constant, mean density of the Universe
and cosmological constant), $k$ and $I$ are the Boltzmann constant and the
mechanical equivalent of heat which play the role of conversion factors
between temperature on the one hand, energy and mechanical units on
the other. After adoption in 1983 of a new definition of the meter
($\lambda = ct$ or $\ell = ct$) this role is partially played also by
the speed of light $c$. It is now also a conversion factor between units of
time (frequency) and length, it is defined with the absolute (null)
accuracy.

Now, when the theory of electroweak interactions has a firm experimental
basis and we have some good models of strong interactions, a more
prefarable choice is as follows:
\beq
\hbar,\ (c),\ e,\ m_e,\ \theta_w,\ G_F,\ \theta_c,\ \Lambda_{QCD},\ G,\ H,\
\rho,\ \Lambda,\ k,\ I
\eeq
and, possibly, three angles of Kobayashi-Maskawa --- $\theta_2$, $\theta_3$
and $\delta$. Here $\theta_w$ is the Weinberg angle, $\theta_c$ is the
Cabibbo angle and $\Lambda_{QCD}$ is a cut-off parameter of quantum
chromodynamics. Of course, if a theory of four known now interactions
will be created (M-, F-or other), then we will probably have another
choice. As we see, the macro constants remain the same, though in some
unified models, i.e. in multidimensional ones, they may be related in some
manner (see below). From the point of view of these unified models the
above mentioned ones are low energy constants.

All these constants are known with different {\em accuracies}. The most
precisely defined constant was and remain the speed of light c: its
accuracy was $10^{-10}$ and now it is defined with the null accuracy.
Atomic
constants, $e$, $\hbar$, $m$ and others are determined with errors
$10^{-6}\div 10^{-8}$, $G$ up to $10^{-4}$ or even worse,
$\theta_w$ --- up to 10\%; the accuracy of $H$ is also about 10\%.
An even worse situation is now with other cosmological parameters (FPC):
mean density estimations vary within an order of magnitude; for $\Lambda$
we have now data that its corresponding density exceeds the matter density
(0.7 of the total mass).

As to the {\em nature\/} of the FPC, we may mention several approaches. One
of the first hypotheses belongs to J.A. Wheeler: in each cycle of the
Universe evolution the FPC arise anew along with physical laws which govern
this evolution.  Thus, the nature of the FPC and physical laws are
connected
with the origin and evolution of our Universe.

A less global approach to the nature of dimensional constants suggests that
they are needed to make physical relations dimensionless or they are
measures of asymptotic states. Really, the speed of light appears in
relativistic theories in factors like $v/c$, at the same time velocities of
usual bodies are smaller than $c$, so it plays also the role of an
asymptotic limit. The same sense have some other FPC: $\hbar$ is the
minimal quantum of action, $e$ is the minimal observable charge (if we do
not take into account quarks which are not observable in a free state) etc.

Finally, FPC or their combinations may be considered as natural scales
determining the basic units. While the earlier basic units were chosen more
or less arbitrarily, i.e., the second, meter and kilogram, now the first
two are based on stable (quantum) phenomena. Their stability is believed to
be ensured by the physical laws which include FPC.

Another interesting problem, which is under discussion, is why the FPC have
values in a very narrow range necessary for supporting life (stability of
atoms, stars lifetime etc.).  There exist several possible but far from
being convincing explanations \cite{B}.  First, that it is a good luck, no
matter how improbable is the set of FPC. Second, that life may exist in
other forms and for another FPC set, of which we do not know. Third, that
all possibilities for FPC sets exist in some universe. And the last but not
the least: that there is some cosmic fine tuning of FPC: some unknown
physical processes bringing FPC to their present values in a long-time
evolution, cycles etc.

An exact knowledge of FPC and precision measurements are necessary for
testing main physical theories, extention of our knowledge of nature and,
in the long run, for practical applications of fundamental theories.
Within this, such theoretical problems arise:

1) development of models for confrontation of theory with experiment in
critical situations (i.e. for verification of GR, QED, QCD, GUT or
other unified models);

2) setting limits for spacial and temporal variations of FPC.

As to a {\em classification\/} of FPC, we may set them now into four groups
according to their generality:

1) Universal constants such as $\hbar$, which divides
all phenomena into quantum and nonquantum ones (micro- and macro-worlds)
and to a certain extent $c$, which divides all motions into relativistic
and non-relativistic ones;

2) constants of interactions like $\alpha$, $\theta_w$, $\Lambda_{QCD}$
and $G$;

3) constants of elementary constituencies of matter like $m_e$, $m_w$,
$m_x$, etc., and

4) transformation multipliers such as $k$, $I$ and partially $c$.

Of course, this division into classes is not absolute. Many
constants move from one class to another. For example, $e$ was a charge
of a particular object -- electron, class 3, then it became a characteristic
of class 2 (electromagnetic interaction, $\alpha=\frac{e^2}{\hbar c}$ in
combination with $\hbar$ and $c$); the speed of light $c$ has been in
nearly all classes: from 3 it moved into 1, then also into 4. Some of
the constants ceased to be fundamental (i.e. densities, magnetic moments,
etc.) as they are calculated via other FPC.

As to the {\em number} of FPC, there are two opposite tendencies: the number
of ``old'' FPC is usually diminishing when a new, more general theory is
created, but at the same time new fields of science arise, new processes
are discovered in which new constants appear. So, in the long run we may
come to some minimal choice which is characterized by one or several FPC,
maybe connected with the so-called Planck parameters --- combinations of
$c$, $\hbar$ and $G$:
\bear
L=\left(\frac{\hbar G}{c^3}\right)^{1/2} \lal\sim 10^{-33}\ {\rm cm}, \nn
m_L = (c\hbar/2G)^{1/2} \lal \sim 10^{-5}\ {\rm g}, \nn
	    \tau_L = L/c \lal\sim 10^{-43}\ {\rm s}.
\ear

The role of these parameters is important since $m_L$ characterizes the
energy of unification of four known fundamental interactions: strong,
weak, electromagnetic and gravitational ones, and $L$ is a scale where
the classical notions of space-time loose their meaning.

\medskip\noi
{\bf 2.2.} The problem of the gravitational constant $G$ measurement and
its stability is a part of a rapidly developing field, called
gravitational-relativistic metrology (GRM). It has appeared due to the
growth of
measurement technology precision, spread of measurements over large scales
and a tendency to the unification of fundamental physical interaction
\cite{6}, where main problems arise and are concentrated on
the gravitational interaction.

The main subjects of GRM are:

- general relativistic models for different astronomical scales: Earth,
Solar System, galaxes, cluster of galaxies, cosmology - for time transfer,
VLBI, space dynamics, relativistic astrometry etc.(pioneering works were
done in Russia by Arifov and Kadyev, Brumberg in 60's);

- development of generalized gravitational theories and unified models for
testing their effects in experiments;

- fundamental physical constants, G in particular, and their stability in
space and time;

- fundamental cosmological parameters as fundamental constants:
cosmological models studies, measurements and observations;

- gravitational waves (detectors, sources...);

- basic standards (clocks) and other modern precision devices (atomic and
neutron
interferometry, atomic force spectroscopy etc.) in fundamental
gravitational experiments, especially in space...

There are three problems related to $G$, which origin lies mainly in
unified models predictions:

1) absolute $G$ measurements,

2) possible time variations of $G$,

3) possible range variations of $G$ -- non-Newtonian, or new interactions.

{\em Absolute measurements of $G$}. There are many laboratory
determinations of $G$ with errors of the order $10^{-3}$ and only 4 on the
level of $10^{-4}$. They are (in $10^{-11}$ \Gun):

\medskip
\begin{tabular}{lll}
1. Facy and Pontikis, France & 1972 --- & 6,6714 $\pm$ 0.0006\\
2. Sagitov et al., Russia    & 1979 --- & 6,6745 $\pm$ 0.0008\\
3. Luther and Towler, USA    & 1982 --- & 6,6726 $\pm$ 0.0005\\
4. Karagioz, Russia          & 1988 --- & 6,6731 $\pm$ 0.0004\\
\end{tabular}
\medskip

>From this table it is evident that the first three experiments contradict
each other (the results do not overlap within their accuracies). And only
the fourth experiment is in accord with the third one.

The official CODATA value of 1986
\beq
G = (6,67259 \pm 0.00085)\cdot 10^{-11}\cdot m^3\cdot kg^{-1}\cdot s^{-2}
\eeq
is based on the Luther and Towler determination. But after very precise
measurements of $G$ in Germany and New Zealand the situation became more
vague. Their results deviate from the official CODATA value by more than
600 ppm.

As it may be seen from the Cavendish conference data \cite{MT99}, the
results of 7 groups may agree with each other only on the level $10^{-3}$.
The most recent and precise $G$-measurement \cite{G} diverge also from the
CODATA value of 1986.

This means that either the limit of terrestrial accuracies has been reached
or we have some new physics entering the measurement proceedure \cite{6}.
The first means that, maybe we should turn to space experiments to
measure
$G$ \cite{15}, and and second means that a more thorough study of theories
generalizing Einstein's general relativity or unified theories is necessary.

There exist also some satellite determinations of $G$ (namely $G\cdot
M_{\rm Earth}$) on the level of $10^{-9}$ and several less
precise geophysical determinations in mines.

The precise knowledge of $G$ is necessary, first of all, as it is a FPC;
next, for the evaluation of mass of the Earth, of planets, their mean
density
and, finally, for construction of Earth models; for transition from
mechanical to electromagnetic units and back; for evaluation of other
constants through relations between them given by unified theories; for
finding new possible types of interactions and geophysical effects; for
some practical applications like increasing of modern gradiometers
precision, as they demand a calibration by a gravitational field of a
standard body depending on G: high accuracy of their calibration
($10^{-5}$ - $10^{-6}$) requires the same accuracy of $G$.
(I am indebted to Dr.N.Kolosnitsyn for this last remark.)

The knowledge of constants values has not only a fundamental meaning but
also a metrological one. The modern system of standards is based mainly on
stable physical phenomena. So, the stability of constants plays a crucial
role. As all physical laws were established and tested during the last 2-3
centuries in experiments on the Earth and in the near space, i.e. at a
rather short space and time intervals in comparison with the radius and
age of the Universe, the possibility of slow {\em variations} of constants
(i.e. with the rate of the evolution of the Universe or slower) cannot be
excluded a priori.

So, the assumption of absolute stability of constants is an
extrapolation and each time we must test it.

{\bf 2.3.} {\em Time Variations of $G$}.
The problem of variations of FPC arose
with the attempts to explain the relations between micro- and macro-world
phenomena. Dirac was the first to introduce (1937) the so-called
``Large Numbers Hypothesis'' which relates some known very big (or very
small) numbers with the dimensionless age of the Universe $T\sim 10^{40}$
(age of the Universe in seconds $10^{17}$, divided by the characteristic
elementary particle time $10^{-23}$ seconds). He suggested (after Milne in
1935) that the ratio of the gravitational to strong interaction
strengths, $Gm_p^2/\hbar c\sim 10^{-40}$, is inversely proportional to the
age of the Universe: $Gm_p^2/\hbar c\sim T^{-1}$. Then, as the age varies,
some constants or their combinations must vary as well. Atomic constants
seemed to Dirac to be more stable, so he chose the variation of $G$ as
$T^{-1}$.

After the original {\it Dirac hypothesis} some new ones appeared
and also some generalized {\em theories} of gravitation admitting the
variations of an effective gravitational coupling. We may single out three
stages in the development of this field:
\begin{enumerate}
\item Study of theories and hypotheses with variations of FPC,
	their predictions and confrontation with experiments (1937-1977).

\item
Creation of theories admitting variations of an effective gravitational
constant in a particular system of units, analyses of experimental and
observational data within these theories \cite{1,3}
(1977-present).

\item
Analyses of FPC variations within unified models \cite{6,4,Mel2}
(present).
\end{enumerate}

Within the development of the first stage from the analysis of the whole
set of existed astronomical, astrophysical, geophysical and laboratory
data, a conclusion was made \cite{1,2} that variations of atomic constants
are excluded, but variations of the effective gravitational constant in
the atomic system of units do not contradict the available experimental
data on the level $10^{-11}\div 10^{-12} \year^{-1}$. Moreover, in
\cite{7,1,2} the conception was worked out that variations of constants are
not absolute but depend on the system of measurements (choice of standards,
units and devices using this or that fundamental interaction). Each
fundamental interaction through dynamics, described by the corresponding
theory, defines the system of units and the system of basic standards.

Earlier reviews of some hypotheses on variations of FPC and experimental
tests can be found in \cite{3,4}.

Following Dyson (1972), we can introduce dimensionless combinations of
micro- and macro-constants:
\bear
\begin{array}{ll}
\alpha = e^2/\hbar c = 7,3\cdot10^{-3}, \cm &\gamma = Gm^2/\hbar c =
	5\cdot10^{-39}, \yy
\beta = G_F\ m^2c/\hbar ^3 = 9\cdot10^6,
	\cm & \delta = H\hbar /mc^2 = 10^{-42},			\yy
\varepsilon = \rho G/H^2 = 2\cdot10^{-3}, & t = T/(e^2/mc^3) \approx 10^{40}
\end{array}
\earn
We see that $\alpha$, $\beta$ and $\varepsilon$ are of order 1 and
$\gamma$ and $\delta$ are of the order $10^{-40}$. Nearly all existing
hypotheses on variations of FPC may be represented as follows:

\noindent
{\bf Hyposesis 1 (standard)}:

$\alpha$, $\beta$, $\gamma$ are constant, $\delta\sim t^{-1}$,
$\varepsilon \sim t$.

Here we have no variations of $G$ while $\delta$ and $\varepsilon$ are
determined by cosmological solutions.

\noindent
{\bf Hyposesis 2 (Dirac)}:

$\alpha$, $\beta$, $\varepsilon$ are constant, $\gamma \sim t^{-1}$,
$\delta \sim t^{-1}$.

Then $\dot G/G = 5\cdot10^{-11} \year^{-1}$ if the age of the
Universe is taken to be $T = 2\ten{10}$ years.

\noindent
{\bf Hyposesis 3 (Gamow)}:

$\gamma /\alpha = Gm^2/e^2 \sim 10^{-37}$, so $e^2$ or $\alpha$ are
varied, but not $G$, $\beta$, $\gamma$; $\varepsilon = \const$, $\alpha
\sim t^{-1}$, $\delta \sim t^{-1}$.

Then $\dot{\alpha}/\alpha = 10^{-10}\ \year^{-1}$.

\noindent
{\bf Hyposesis 4 (Teller)}:

trying to account also for deviations of $\alpha$ from 1, he suggested
$\alpha^{-1} = \ln\gamma^{-1}$.

Then $\beta$, $\varepsilon$ are constants, $\gamma \sim t^{-1}$, $\alpha
\sim (\ln t)^{-1}$, $\delta \sim t^{-1}$
\beq
	\dot{\alpha}/\alpha = 5\cdot10^{-13} year^{-1}.
\eeq
The same relation for $\alpha$ and $\gamma$ was used also by Landau, DeWitt,
Staniukovich, Terasawa and others, but in approaches other than Teller's.

Some other variants may be also possible, e.g. the Brans-Dicke theory with
$G \sim t^{-r}$, $\rho \sim t^{r-2}$, $r=[2+3\omega/2]^{-1}$, a
combination of Gamow's and Brans-Dicke etc. \cite{3}.

\medskip\noi
{\bf 2.4.} There are different astronomical, geophysical and laboratory
{\em data} on possible variations of FPC.

\medskip\noi
{\bf Astrophysical data}:
\begin{description}
\item[a)] from comparison of fine structure $(\sim \alpha^2)$ and
relativistic fine structure $(\sim \alpha^4)$ shifts in spectra of
radio galaxies, Bahcall and Schmidt (1967) obtained
\beq
|\dot {\alpha}/\alpha| \leq 2\cdot10^{-12}\ \year^{-1};
\eeq

\item[b)] comparing lines in optical $(\sim Ry=me^4/\hbar ^2)$ and radio
bands of the same sources in galaxies Baum and Florentin-Nielsen (1976)
got the estimate
\beq
	| \dot {\alpha}/\alpha | \leq 10^{-13} \ \year^{-1},
\eeq
and for extragalactic objects
\beq
	| \dot \alpha /\alpha | \leq 10^{-14} \ \year^{-1};
\eeq

\item[c)] from observations of superfine structure in H-absorption lines
of the distant radiosource Wolf et al. (1976) obtained that
\beq
	| \alpha^2 (m_e/m_p) g_p| < 2\cdot10^{-14}\ ;
\eeq
\end{description}

>from these data it is seen that Hyposeses 3 and 4 are excluded.  Recent
data only strengthen this conclusion. Comparing the data from absorption
lines of atomic and molecular transition spectra in high redshifts QSO's,
Varshalovich and Potekhin, Russia, \cite{VP} obtained for $z = 2.8 - 3.1$:
\beq
	| \dot \alpha /\alpha | \leq 1.6\cdot10^{-14} \ \year^{-1}
\eeq
and Drinkwater et al. \cite{D}:
\beq
	| \dot \alpha /\alpha | \leq 10^{-15} \ \year^{-1}
		\mbox{ for } z = 0.25
\eeq
and
\beq
	| \dot \alpha /\alpha | \leq 5\cdot 10^{-16} \ \year^{-1}
	\mbox{ for } z = 0.68
\eeq
for a model with zero deceleration parameter and $H = 75 {\rm km}\cdot
{\rm s}^{-1}\cdot {\rm Mpc}^{-1}$.

The same conclusion is made on the bases of {\em geophysical data}. Indeed,
\begin{description}
\item[a)]
$\alpha$-decay of $U_{238}\to Pb_{208}$. Knowing abundancies of
$U_{238}$ and $P_{238}$ in rocks and independently the age of these rocks,
one obtains the limit
\beq
	| \dot {\alpha}/\alpha | \leq 2\cdot10^{-13} \ \year^{-1};
\eeq
\item
[b)] from spontaneous fission of $U_{238}$ such an estimation was made:
\beq
	| \dot {\alpha}/\alpha | \leq 2,3\cdot10^{-13} \ \year^{-1}.
\eeq
\item
[c)] finally, $\beta$-decay of $Re_{187}$ to $Os_{187}$ gave:
\beq
	| \dot {\alpha}/\alpha | \leq 5\cdot10^{-15} \ \year^{-1}
\eeq
\end{description}

We must point out that all astronomical and geophysical estimations are
strongly model-dependent. So, of course, it is always desirable to have
{\em laboratory tests\/} of variations of FPC.
\begin{description}
\item
[a)] Such a test was first done by the Russian group in the Committee for
Standards (Kolosnitsyn, 1975). Comparing rates of two different types of
clocks, one based on a Cs standard and another on a beam molecular
generator, they found that
\beq
	| \dot {\alpha}/\alpha | \leq 10^{-10} \ \year^{-1}.
\eeq

\item
[b)] From a similar comparison of a Cs standard and SCCG (Super Conducting
Cavity Generator) clocks rates Turneaure et al. (1976) obtained the limit
\beq
	| \dot {\alpha}/\alpha | \leq 4.1\cdot 10^{-12} \ \year^{-1}.
\eeq

\item
[c)] More recent data were obtained by J. Prestage et al. \cite{JP}
by comparing mercury and $H$-maser clocks. Their result is
\beq
	| \dot {\alpha}/\alpha | \leq 3.7\cdot 10^{-14} \ \year^{-1}.
\eeq
All these limits were placed on the fine structure constant variations. From
the analysis of decay rates of $K_{40}$ and $Re_{187}$, a limit on
possible variations of the weak interaction constant was obtained (see
approach for variations of $\beta$, e.g. in \cite{8})
\beq
	| \dot {\beta}/\beta | \leq 10^{-10} \ \year^{-1}.
\eeq
But the most strict data were obtained by A. Schlyachter in 1976 (Russia) 
>from an
analysis of the ancient natural nuclear reactor data in Gabon, Oklo, because
the event took place $2\cdot10^9$ years ago. They are the following:
\bear
| \dot {G}_s/G_s| < 5\cdot10^{-19} \ \year^{-1}, \nonumber \\
| \dot {\alpha}/\alpha | < 10^{-17} \ \year^{-1}, \\
| \dot {G}_F/G_F| < 2\cdot10^{-12} \ \year^{-1}. \nonumber
\ear
Quite recently Damour and Dyson \cite{13} repeated this analysis in more
detail and gave more cautious results:
\beq
| \dot {\alpha}/\alpha | \leq 5\cdot 10^{-17} \ \year{-1}
\eeq
and
\beq
| \dot {G}_F/G_F| < 10^{-11} \ \year^{-1}.
\eeq
So, we really see that all existing hyposeses with variations of atomic
constants are excluded.
\end{description}
\noi
{\bf 2.5.} Now we still have no unified theory of all four interactions.
So it is possible to construct systems of measurements based on any of
these four interactions. But practically it is done now on the basis of
the mostly worked out theory --- on electrodynamics (more precisely on
QED). Of course, it may be done also on the basis of the gravitational
interaction (as it was partially earlier). Then, different units of basic
physical quantities arise based on dynamics of the given interaction, i.e.
the atomic (electromagnetic) second, defined via frequency of atomic
transitions or the gravitational second defined by the mean Earth motion
around the Sun (ephemeris time).

It does not follow from anything that these two seconds are always
synchronized in time and space. So, in principal they may evolve
relative to each other, for example at the rate of the evolution of the
Universe or at some slower rate.

That is why, in general, variations of the gravitational constant are
possible in the atomic system of units ($c$, $\hbar$, $m$ are constant) and
masses of all particles --- in the gravitational system of units ($G$,
$\hbar$, $c$ are constant by definition).  Practically we can test only
the first variant since the modern basic standards are defined in the
atomic system of measurements. Possible variations of FPC must be tested
experimentally but for this it is necessary to have the corresponding
theories admitting such variations and their certain effects.

Mathematically these systems of measurement may be realized as
conformally related metric forms. Arbitrary conformal transformations give
us a transition to an arbitrary system of measurements.

We know that scalar-tensor and multidimensional theories are corresponding
frameworks for these variations. So, one of the ways to describe variable
gravitational coupling is the introduction of a {\em scalar field} as an
additional variable of the gravitational interaction. It may be done by
different means (e.g. Jordan, Brans-Dicke, Canuto and others). We have
suggested a variant of gravitational theory with a conformal scalar field
(Higgs-type field \cite{9,3}) where Einstein's general relativity may be
considered as a result of spontaneous symmetry breaking of conformal
symmetry (Domokos, 1976) \cite{3}. In our variant spontaneous symmetry
breaking of the global gauge invariance leads to a nonsingular cosmology
\cite{16}.  Besides, we may get variations of the effective gravitational
constant in the atomic system of units when $m$, $c$, $\hbar$ are constant
and variations of all masses in the gravitational system of units ($G$,
$c$, $\hbar$ are constant). It is done on the basis of approximate
\cite{10} and exact cosmological solutions with local inhomogenity
\cite{14}.

The effective gravitational constant is calculated using the equations of
motions. Post-Newtonian expansion is also used in order to confront the
theory with existing experimental data. Among the post-Newtonian parameters
the parameter $f$ describing variations of $G$ is included. It is defined
as
\beq
	\frac{1}{GM}\frac{d(GM)}{dt} = fH.
\eeq
According to Hellings' data \cite{18} from the Viking mission,
\beq
	\tilde{\gamma}-1 = (-1.2\pm 1.6)\cdot 10^3, \cm
		f = (4\pm 8)\cdot 10^{-2}.
\eeq
In the theory with a conformal Higgs field \cite{10,14} we obtained the
following relation between $f$ and $\tilde\gamma$:
\beq
	f = 4(\tilde{\gamma}-1).
\eeq
Using Hellings' data for $\tilde{\gamma}$, we can calculate in our
variant $f$ and compare it with $f$ from \cite{18}. Then we get
$f=(-9,6\pm 12,8)\cdot 10^{-3}$ which agrees with (24) within its
accuracy.

We used here only Hellings' data on variations of $G$. But the situation
with experiment and observations is not so simple. Along with \cite{18},
there are some other data \cite{3,4}:
\begin{enumerate}
\item
>From the growth of corals, pulsar spin down, etc. on the level
\beq
	| \dot {G}/G| < 10^{-10}\div 10^{-11} \ \year^{-1}.
\eeq

\item
Van Flandern's positive data (though not confirmed and critisized) from
the analysis of
lunar mean motion around the Earth and ancient eclipses data (1976, 1981):
\beq
	| \dot {G}/G| = (6\pm 2)10^{-11}\ \year^{-1}.
\eeq

\item
Reasenberg's estimates (1987) of the same Viking mission as in \cite{18}:
\beq
	| \dot {G}/G| < (0\pm 2)\cdot 10^{-11}\ \year^{-1}
\eeq

\item
Hellings' result in the same form is
\beq
	| \dot {G}/G| < (2\pm 4)\cdot 10^{-12}\ \year^{-1}.
\eeq

\item
A result from nucleosythesis (Acceta et al., 1992):
\beq
	|\dot{G}/G|<(\pm 0.9)\cdot 10^{-12} year^{-1}.
\eeq

\item
E.V.Pitjeva's result, Russia (1997), based on satellites and planets
motion:
\beq
        |\dot{G}/G|<(0\pm 2)\cdot 10^{-12} year^{-1}
\eeq
\end{enumerate}

As we see, there is a vivid contradiction in these results. As to other
experimental or observational data, the results are rather inconclusive.
The most reliable ones are based on lunar laser ranging (Muller et al, 1993
and Williams et al, 1996).  They are not better than $10^{-12}$ per year.
Here, once more we see that there is a need for corresponding theoretical
and experinmental studies. Probably, future space missions like Earth
SEE-satellite \cite{15} or missions to other planets
and lunar laser ranging will be a decisive step in solving the problem of
temporal variations of $G$ and determining the fates of different theories
which predict them, since the greater is the time interval between
successive masurements and, of course, the more precise they are, the more
stringent results will be obtained.

As we saw, different theoretical schemes lead to temporal variations of the
effective gravitational constant:
\begin{enumerate}
\item
Empirical models and theories of Dirac type, where $G$ is replaced by
$G(t)$.

\item
Numerous scalar-tensor theories of Jordan-Brans-Dicke type where $G$
depending on the scalar field $\sigma (t)$ appears.

\item
Gravitational theories with a conformal scalar field arising in different
approaches \cite{3,20}.

\item
Multidimensional unified theories in which there are dilaton fields and
effective scalar fields appearing in our 4-dimensional spacetime from
additional dimensions \cite{19,Mel2}. They may help also in solving the
problem of a variable cosmological constant from Planckian to present
values.
\end{enumerate}

As was shown in \cite{4,19,Mel2} temporal variations of FPC are
connected with each other in {\em multidimensional models} of unification
of interactions. So, experimental tests on $\dot {\alpha}/\alpha$
may at the same time be used for estimation of $\dot {G}/G$ and
vice versa. Moreover, variations of $G$ are related also to the
cosmological parameters $\rho$, $\Omega$ and $q$ which gives opportunities
of raising the precision of their determination.

As variations of FPC are closely connected with the behaviour of internal
scale factors, it is a direct probe of properties of extra dimensions and
the corresponding theories \cite{IM1,BIM1,Mel2}.

\medskip\noi
{\bf 2.6.} {\em Non-Newtonian interactions, or range variations of $G$}.
Nearly all modified theories of gravity and unified theories predict also
some deviations from the Newton law (inverse square law, ISL) or
composition-dependent violations of the Equivalence Principle (EP) due to
appearance of new possible massive particles (partners) \cite{4}.
Experimental data exclude the existence of these particles at nearly all
ranges except less than {\it millimeter\/} and also at {\em meters and
hundreds of meters} ranges. The most recent result in the range of 20 to
500 m was obtained by Achilli et al. using an energy storage plant
experiment with gravimeters. They found a positive result for the
deviation from the Newton law with the Yukawa potential strength $\alpha$
between 0.13 and 0.25. Of course, these results need to be verified in
other independent experiments, probably in space ones \cite{15}.

In the Einstein theory $G$ is a true constant. But, if we think
that $G$ may vary with time, then, from a relativistic point of view, it may
vary with distance as well.  In GR massless gravitons are mediators of the
gravitational interaction, they obey second-order differential equations
and interact with matter with a constant strength $G$. If any of these
requirements is violated, we come in general to deviations from
the Newton law with range (or to generalization of GR).

In \cite{5} we analyzed several classes of such theories:

1. Theories with massive gravitons like bimetric ones or theories with
a $\Lambda$-term.

2. Theories with an effective gravitational constant like the general
scalar-tensor ones.

3. Theories with torsion.

4. Theories with higher derivatives (4th-order equations etc.), where
massive modes appear leading to short-range additional forces.

5. More elaborated theories with other mediators besides gravitons
(partners), like supergravity, superstrings, M-theory etc.

6. Theories with nonlinearities induced by any known physical interactions
(Born-Infeld etc.)

7. Phenomenological models where the detailed mechanism of deviation is
not known (fifth or other force).

In all these theories some effective or real masses appear leading to
Yukawa-type deviation from the Newton law, characterized by strength and
range.

There exist some model-dependant estimations of these forces.  The most
well-known one belongs to Scherk (1979) from supergravity where the
graviton is accompanied by a spin-1 partner (graviphoton) leading to
an additional repulsion.  Other models were suggested by Moody and Wilczek
(1984) -- introduction of a pseudo-scalar particle -- leading to
an additional attraction between macro-bodies with the range
$2\cdot10^{-4}$ cm $< \lambda < 20$ cm and strength $\alpha$ from 1 to
$10^{-10}$ in this range.  Another supersymmetric model was elaborated by
Fayet (1986, 1990), where a spin-1 partner of a massive graviton gives
an additional repulsion in the range of the order $10^{3}$ km and $\alpha$
of the order $10^{-13}$.

A scalar field to adjust $\Lambda$ was introduced also by S. Weinberg in
1989, with a mass smaller than $10^{-3} {\rm eV}/c^{2}$, or a range greater
than 0.1 mm.  One more variant was suggested by Peccei, Sola and Wetterich
(1987) leading to additional attraction with a range smaller than 10 km.
Some $p$-brane models also predict non-Newtonian additional interactions in
the mm range, what is intensively discussed nowadays. About PPN parameters
for multidimensional models with $p$-branes see below.

\medskip\noi
{\bf 2.7. SEE - Project}

We saw that there are three problems connected with $G$. There is a
promising new multi-purpose space experiment SEE - Satellite Energy
Exchange \cite{15}, which addresses all these problems and may be more
effective in solving them than other laboratory or space experiments.

This experiment is based on a limited 3-body problem of celestial
mechanics: small and large masses in a drag-free satellite and the Earth.
Unique horse-shoe orbits, which are effectively one-dimensional, are used
in it.

The aims of the SEE-project are to measure: Inverse Square law (ISL) and
Equivalence Principle (EP) at ranges of meters and the Earth radius,
$G$-dot
and the absolute value of $G$ with unprecedented accuracies.

We studied some aspects of the SEE-project \cite{Team} :

1. Wide range of trajectories with the aim of finding optimal ones:

- circular in spherical field;

- the same plus Earth quadrupole modes;

- elliptic with eccentricity less than 0.05.

2. Estimations of other celestial bodies influence.

3. Estimation of relative influence of trajectories to changes in $G$
and $\alpha$.

4. Modelling measurement procedures for $G$ and $\alpha$ by different
methods, for different ranges and for different satellite altitudes:
optimal - 1500 $km$, ISS free flying platform - 500 $km$ and also for
3000 $km$.

5. Estimations of some sources of errors:

- radial oscillations of the shepherd's surface;

- longitudal oscillations of the capsule;

- transversal oscillations of the calsule;

- shepherd's nonsphericity;

- limits on the quadrupole moment of the shepherd;

- limits on addmissible charges and time scales of charging by
high energy particles etc.

6. Error budgets for $G$, $G$-dot and $G(r)$.

The general conclusion is that the SEE-project may really improve our
knowledge of these values by 3-4 orders  better than we have nowadays.

\section{Multidimensional Models}

The history of the multidimensional approach begins with the well-known
papers of T.K. Kaluza and O. Klein on 5-dimensional theories which opened
an interest to investigations in multidimensional gravity. These ideas were
continued by P. Jordan who suggested to consider the more general case
$g_{55}\ne{\rm const}$ leading to a theory with an additional scalar
field. They were in some sense a source of inspiration for C. Brans
and R.H. Dicke in their well-known work on a scalar-tensor gravitational
theory. After their work a lot of investigations have been performed using
material or fundamental scalar fields, both conformal and non-conformal
(see details in \cite{3}).

A revival of ideas of many dimensions started in the 70's and continues
now.
It is completely due to the development of unified theories. In the 70's
an interest to multidimensional gravitational models was stimulated mainly
by (i) the ideas of gauge theories leading to a non-Abelian
generalization of the Kaluza-Klein approach and (ii) by supergravitational
theories. In the 80's the supergravitational theories were ``replaced" by
superstring models. Now it is heated by expectations connected with
the overall M-theory. In all these theories,
4-dimensional gravitational models with extra fields were obtained from
some multidimensional model by dimensional reduction based on the
decomposition of the manifold
\beq
	M=M^4\times M_{\rm int},
\eeq
where $M^4$ is a 4-dimensional manifold and $M_{\rm int}$ is some internal
manifold (mostly considered to be compact).

The earlier papers on multidimensional gravity and cosmology dealt with
multidimensional Einstein equations and with a block-diagonal cosmological
or spherically symmetric metric defined on the manifold $M= \R \times
M_0\times \dots \times M_n$ of the form
\beq
	g=-dt\otimes dt+\sum_{r=0}^n a_r^2(t) g^r
\eeq
where $(M_r,g^r)$ are Einstein spaces, $r=0,\dots,n$.  In some of them a
cosmological constant and simple scalar fields were also used \cite{BIMZ}.

Such models are usually reduced to pseudo-Euclidean Toda-like systems with
the Lagrangian
\beq
L=\frac12G_{ij}\dot x^i\dot x^j-\sum_{k=1}^mA_k{\rm e}^{u_i^kx^i}
\eeq
and the zero-energy constraint $E=0$.

It should be noted that pseudo-Euclidean Toda-like systems are not
well-studied yet. There exists a special class of equations of state that
gives rise to Euclidean Toda models \cite{GIM}.

Cosmological solutions are closely related to solutions with spherical
symmetry \cite{IME}. Moreover, the scheme of obtaining the latter is very
similar to the cosmological approach \cite{Mel2}. The first
multidimensional generalization of such type was considered by D. Kramer
and rediscovered by A.I. Legkii, D.J.  Gross and M.J. Perry (and also by
Davidson and Owen). In \cite{BrI} the Schwarzschild solution was
generalized to the case of $n$ internal Ricci-flat spaces and it was shown
that a black hole configuration takes place when the scale factors of
internal spaces are constants. It was shown there also that a minimally
coupled scalar field is incompatible with the existence of black holes. In
\cite{FIM2} an analogous generalization of the Tangherlini solution was
obtained, and an investigation of singularities was performed in
\cite{IMB}.  These solutions were also generalized to the electrovacuum
case with and without a scalar field \cite{FIM3,IM8,BM}. Here, it was
also proved that BHs exist only when a scalar field is switched
off. Deviations from the Newton and Coulomb laws were obtained depending on
mass, charge and number of dimensions. In \cite{BM} spherically symmetric
solutions were obtained for a system of scalar and electromagnetic fields
with a dilaton-type interaction and also deviations from the Coulomb law
were calculated depending on charge, mass, number of dimensions and
dilaton coupling.  Multidimensional dilatonic black holes were singled out.
A theorem was proved in \cite{BM} that ``cuts" all non-black-hole
configurations as being unstable under even monopole perturbations. In
\cite{IM13} the extremely charged dilatonic black hole solution was
generalized to a multicenter (Majumdar-Papapetrou) case when the
cosmological constant is non-zero.

We note that for $D =4$ the pioneering Majumdar-Papapetrou solutions with
a conformal scalar field and an electromagnetic field were considered in
\cite{Br}.

At present there exists a special interest to the so-called M- and
F-theories etc. These theories are ``supermembrane" analogues of
the superstring models in $D=11,12$ etc. The low-energy limit of these
theories leads to models governed by the Lagrangian
\beq
{\cal L} = R[g]-
h_{\alpha\beta} g^{MN}\d_{M}\varphi^\alpha\d_{N}\varphi^\beta
-\sum_{a\in\Delta}\frac{\theta_a}{n_a!}\exp[2\lambda_{a}(\varphi)](F^a)^2,
\eeq
where $g$ is a metric, $F^a=dA^a$ are forms of rank $F^a=n_a$, and
$\varphi^\alpha$ are scalar fields.

In \cite{IMC} it was shown that, after dimensional reduction on the
manifold $M_0\times M_1\times\dots\times M_n$ and when the composite
$p$-brane ansatz is considered, the problem is reduced to the gravitating
self-interacting $\sigma$-model with certain constraints. For
electric $p$-branes see also \cite{IM0,IM,IMR} (in \cite{IMR} the
composite electric case was considered). This representation may be
considered as a powerful tool for obtaining different solutions with
intersecting $p$-branes (analogs of membranes).  In \cite{IMC,IMBl}
Majumdar-Papapetrou type solutions were obtained (for the non-composite
electric case see \cite{IM0,IM} and for the composite electric case see
\cite{IMR}). These solutions correspond to Ricci-flat $(M_i,g^i)$,
$i=1,\dots,n$ and were generalized to the case of Einstein internal
spaces \cite{IMC}.  The obtained solutions take place when certain
{\em orthogonality relations} (on couplings parameters, dimensions of
``branes", total dimension) are imposed. In this situation a class of
cosmological and spherically symmetric solutions was obtained \cite{IMJ}.
Special cases were also considered in \cite{BKR}. Solutions with a
horizon were considered in detail in \cite{BIM,IMJ}. In \cite{BIM,Br1}
some propositions related to (i) interconnection between the Hawking
temperature and the singularity behaviour, and (ii) to multitemporal
configurations were proved.

It should be noted that multidimensional and multitemporal generalizations
of the Schwarz\-schild and Tangherlini solutions were considered in
\cite{IM8,IM6I}, where the generalized Newton formulas in a multitemporal
case were obtained.

We note also that  there exists a large variety of Toda solutions (open or
closed) when certain intersection rules are satisfied \cite{IMJ}.

We continued our investigations of $p$-brane solutions
based on the sigma-model approach in \cite{IMC,IM,IMR}. (For the pure
gravitational sector see \cite{IM0,S,IMC}.)

We found a family of solutions depending on one variable describing
the (cosmological or spherically symmetric) ``evolution" of $(n+1)$
Einstein spaces in the theory with several scalar fields and forms. When
an electro-magnetic composite $p$-brane ansatz is adopted, the field
equations are reduced to the equations for a Toda-like system.

In the case when $n$ ``internal" spaces are Ricci-flat, one space $M_0$ has
a non-zero curvature, and all $p$-branes do not ``live" in $M_0$, we found a
family of solutions to the equations of motion (equivalent to
equations for Toda-like Lagrangian with zero-energy constraint \cite{IMJ})
if certain {\em block-orthogonality relations} on $p$-brane vectors $U^s$
are imposed. These solutions generalize the solutions from \cite{IMJ} with
an orthogonal set of vectors $U^s$. A special class of ``block-orthogonal"
solutions (with coinciding parameters $\nu_s$ inside blocks) was
considered earlier in \cite{Br1}.

We considered a subclass of spherically symmetric solutions.  This
subclass contains non-extremal $p$-brane black holes for zero values of
``Kasner-like" parameters. A relation for the Hawking temperature was
presented (in the black hole case).

We also calculated the Post-Newtonian Parameters $\beta$ and $\gamma$
(Eddington parameters) for general spherically symmetric solutions
and black holes in particular \cite{Spr}. These
parameters depending on $p$-brane charges, their worldvolume dimensions,
dilaton couplings and number of dimensions may be useful for possible
physical applications.

Some specific models in classical and quantum multidimensional cases with
$p$-branes were analysed in \cite{GIMT}. Exact solutions for the system of
scalar fields and fields of forms with a dilatinic type interactions for
{\em generalized intersection rules} were studied in \cite{IMOP}, where the
PPN parameters were also calculated.

Finally, a {\em stability\/} analysis for solutions with $p$-branes was
carried out \cite{STAB}. It was shown there that for some simple $p$-brane
systems multidimensional black branes are stable under monopole
perturbations while other (non-BH) sperically symmetric solutions turned
out to be unstable.

\Acknow
{The author is very grateful to Don Strayer, Ulf
Israelsson and Monica
King for their hospitality during his stay in Solvang. The work was
supported in part by the NASA/SEE project and CONACYT, Mexico}

\end{document}